\pdfoutput=1

\newcommand\apjcls{1}
\newcommand\aastexcls{2}
\newcommand\othercls{3}

\newcommand\papercls{\aastexcls}
\documentclass[tighten, times, preprint2]{aastex6}  

\if\papercls \apjcls
\usepackage{apjfonts}
\else\if\papercls \othercls
\usepackage{epsfig}
\fi\fi
\usepackage{ifthen}
\usepackage{natbib}
\usepackage{amssymb, amsmath}
\usepackage{appendix}
\usepackage{etoolbox}
\usepackage[T1]{fontenc}
\usepackage{paralist}

\if\papercls \apjcls
\newcommand\aas{\ref@jnl{AAS Meeting Abstracts}}
\newcommand\dps{\ref@jnl{AAS/DPS Meeting Abstracts}}
\newcommand\maps{\ref@jnl{MAPS}}
\else\if\papercls \othercls
\usepackage{astjnlabbrev-jh}
\fi\fi

\if\papercls \aastexcls
\hypersetup{citecolor=blue, 
            linkcolor=blue, 
            menucolor=blue, 
            urlcolor=blue}  
\else
\usepackage[
bookmarks=true,           
bookmarksnumbered=true,   
colorlinks=true,          
citecolor=blue,           
linkcolor=blue,           
menucolor=blue,           
urlcolor=blue,            
linkbordercolor={0 0 1},  
pdfborder={0 0 1},
frenchlinks=true]{hyperref}
\fi

\if\papercls \othercls

\else

\fi

\providecommand{\adsurl}[1]{\href{#1}{ADS}}

\makeatletter
\patchcmd{\NAT@citex}
  {\@citea\NAT@hyper@{%
     \NAT@nmfmt{\NAT@nm}%
     \hyper@natlinkbreak{\NAT@aysep\NAT@spacechar}{\@citeb\@extra@b@citeb}%
     \NAT@date}}
  {\@citea\NAT@nmfmt{\NAT@nm}%
   \NAT@aysep\NAT@spacechar\NAT@hyper@{\NAT@date}}{}{}

\patchcmd{\NAT@citex}
  {\@citea\NAT@hyper@{%
     \NAT@nmfmt{\NAT@nm}%
     \hyper@natlinkbreak{\NAT@spacechar\NAT@@open\if*#1*\else#1\NAT@spacechar\fi}%
       {\@citeb\@extra@b@citeb}%
     \NAT@date}}
  {\@citea\NAT@nmfmt{\NAT@nm}%
   \NAT@spacechar\NAT@@open\if*#1*\else#1\NAT@spacechar\fi\NAT@hyper@{\NAT@date}}
  {}{}
\makeatother

\makeatletter
\DeclareRobustCommand{\lowcase}[1]{\@lowcase#1\@nil}
\def\@lowcase#1\@nil{\if\relax#1\relax\else\MakeLowercase{#1}\fi}
\makeatother

\DeclareSymbolFont{UPM}{U}{eur}{m}{n}
\DeclareMathSymbol{\umu}{0}{UPM}{"16}
\let\oldumu=\umu
\renewcommand\umu{\ifmmode\oldumu\else\math{\oldumu}\fi}

\if\papercls \othercls

\else

\fi

\let\oldsim=\sim
\renewcommand\sim{\ifmmode\oldsim\else\math{\oldsim}\fi}
\let\oldpm=\pm
\renewcommand\pm{\ifmmode\oldpm\else\math{\oldpm}\fi}
\newcommand\by{\ifmmode\times\else\math{\times}\fi}

\newcommand\tablebox[1]{\begin{tabular}[t]{@{}l@{}}#1\end{tabular}}
\newbox{\wdbox}
\renewcommand\c{\setbox\wdbox=\hbox{,}\hspace{\wd\wdbox}}
\renewcommand\i{\setbox\wdbox=\hbox{i}\hspace{\wd\wdbox}}

\newcount\timect
\newcount\hourct
\newcount\minct
\newcommand\now{\timect=\time \divide\timect by 60
         \hourct=\timect \multiply\hourct by 60
         \minct=\time \advance\minct by -\hourct
         \number\timect:\ifnum \minct < 10 0\fi\number\minct}

\catcode`@=11

\newcommand\comment[1]{}

\newcommand\commenton{\catcode`\%=14}

\renewcommand\math[1]{$#1$}
\newcommand\mathshifton{\catcode`\$=3}

\let\atab=&
\newcommand\atabon{\catcode`\&=4}

\let\oldmsp=\sp
\let\oldmsb=\sb
\def\sp#1{\ifmmode
           \oldmsp{#1}%
         \else\strut\raise.85ex\hbox{\scriptsize #1}\fi}
\def\sb#1{\ifmmode
           \oldmsb{#1}%
         \else\strut\raise-.54ex\hbox{\scriptsize #1}\fi}
\newbox\@sp
\newbox\@sb
\def\sbp#1#2{\ifmmode%
           \oldmsb{#1}\oldmsp{#2}%
         \else
           \setbox\@sb=\hbox{\sb{#1}}%
           \setbox\@sp=\hbox{\sp{#2}}%
           \rlap{\copy\@sb}\copy\@sp
           \ifdim \wd\@sb >\wd\@sp
             \hskip -\wd\@sp \hskip \wd\@sb
           \fi
        \fi}
\def\msp#1{\ifmmode
           \oldmsp{#1}
         \else \math{\oldmsp{#1}}\fi}
\def\msb#1{\ifmmode
           \oldmsb{#1}
         \else \math{\oldmsb{#1}}\fi}

\def\supon{\catcode`\^=7}

\def\subon{\catcode`\_=8}

\def\supsubon{\supon \subon}

\newcommand\actcharon{\catcode`\~=13}

\newcommand\paramon{\catcode`\#=6}

\comment{And now to turn us totally on and off...}

\newcommand\reservedcharson{ \commenton  \mathshifton  \atabon  \supsubon 
                             \actcharon  \paramon}

\catcode`@=12
\reservedcharson

\if\papercls \apjcls

\else

\fi

\if\papercls \othercls
\else
  \newcommand\inpress{n}
  \if\inpress y
    \received{\today}
    \revised{}
    \accepted{}
    \if\papercls \apjcls
    \slugcomment{}
    \fi
  \else
  \slugcomment{\tablebox{In preparation for {\em ApJ}. DRAFT of {\today}.}}
  \fi
\fi

\newcommand\chisq{\ifmmode{\chi\sp{2}}\else\math{\chi\sp{2}}\fi}
\newcommand\redchisq{\ifmmode{ \chi\sp{2}\sb{\rm red}}
                    \else\math{\chi\sp{2}\sb{\rm red}}\fi}
\newcommand\Teq{\ifmmode{T\sb{\rm eq}}\else$T$\sb{eq}\fi}
\newcommand\mjup{\ifmmode{M\sb{\rm Jup}}\else$M$\sb{Jup}\fi}
\newcommand\rjup{\ifmmode{R\sb{\rm Jup}}\else$R$\sb{Jup}\fi}
\newcommand\msun{\ifmmode{M\sb{\odot}}\else$M\sb{\odot}$\fi}
\newcommand\rsun{\ifmmode{R\sb{\odot}}\else$R\sb{\odot}$\fi}
\newcommand\mearth{\ifmmode{M\sb{\oplus}}\else$M\sb{\oplus}$\fi}
\newcommand\rearth{\ifmmode{R\sb{\oplus}}\else$R\sb{\oplus}$\fi}

\usepackage{color}

\shorttitle{47 Tuc X9: an accreting black hole}
\shortauthors{Church {\em et al.}}

\begin{document}

\title{Formation constraints indicate a black-hole accretor in 47 Tuc X9}

\author{Ross~P.~Church\altaffilmark{1},
Jay~Strader\altaffilmark{2},
Melvyn~B.~Davies\altaffilmark{1},
and
Alexey~Bobrick\altaffilmark{1}
}

\affil{\sp{1} Lund Observatory, Department of Astronomy and Theoretical Physics, Box 43, SE 221-00 Lund, Sweden\\
       \sp{2} Department of Physics and Astronomy, Michigan State University, East Lansing, MI 48824, USA
              }

\email{ross@astro.lu.se}

\begin{abstract}
   The luminous X-ray binary 47 Tuc X9 shows radio and X-ray emission consistent
   with a stellar-mass black hole accreting from a carbon-oxygen white dwarf.
   Its location, in the core of the massive globular cluster 47 Tuc, hints at a
   dynamical origin.  We assess the stability of mass transfer from a
   carbon-oxygen white dwarf onto compact objects of various masses, and
   conclude that for mass transfer to proceed stably the accretor must, in fact,
   be a black hole.  Such systems can form dynamically by the collision of a
   stellar-mass black hole with a giant star.  Tidal dissipation of
   energy in the giant's envelope leads to a bound binary with a pericentre
   separation less than the radius of the giant.  An episode of common-envelope
   evolution follows, which ejects the giant's envelope. We find that the most
   likely target is a horizontal-branch star, and that a realistic quantity of
   subsequent dynamical hardening is required for the resulting binary to merge
   via gravitational wave emission.  Observing one binary like 47 Tuc X9 in the
   Milky Way globular cluster system is consistent with the
   expected formation rate.  The observed 6.8-day periodicity in the X-ray
   emission may be driven by eccentricity induced in the UCXB's orbit by a
   perturbing companion.
\end{abstract}


\section{INTRODUCTION}
\label{introduction}

X9 is the most luminous (a few $\times\,10^{33}\,{\rm erg\,s}^{-1}$) hard X-ray source
in the massive globular cluster 47 Tuc \citep{Grindlay+01}. With a UV-bright
variable optical counterpart and hints of variability on timescales from minutes
to hours, the source has generally been classified as an accreting white dwarf
\citep{Paresce+92,Edmonds+03,Heinke+05,Knigge+08}. This consensus was
challenged by the detection of persistent radio continuum emission from the
system \citep{Miller-Jones+15}, much brighter than ever detected from a
cataclysmic variable at this X-ray luminosity.  Instead, the ratio of radio to
X-ray flux is consistent with that observed for quiescent stellar-mass black
holes. 

Several other properties of the source are also unusual for an accreting white
dwarf.  The X-ray spectrum shows very
strong emission tentatively identified as \ion{O}{8} \citep{Heinke+05}, and
double-peaked \ion{C}{4} emission was detected in far-UV STIS spectroscopy
\citep{Knigge+08}. This latter line itself is not unusual for white dwarf
accretors, but the separation of the two emission peaks was nearly $\sim 2100$
km s$^{-1}$, with emission extending to at least 4000 km s$^{-1}$ from rest,
consistent with the \ion{C}{4} being produced in an accretion disk around a
compact object more massive than a white dwarf.

\citet{Bahramian+17} present a comprehensive analysis of the voluminous X-ray
data for X9. They find a 28.2-min periodic signal in the X-ray light curves
taken with two independent \emph{Chandra} instruments at different epochs, which
they identify as the likely orbital period for the system. This short period
implies an ultra-compact system with a white dwarf donor, which is
backed up by the lack of evidence for strong hydrogen emission
\citep{Miller-Jones+15}.  The X-ray spectrum is best fit by photoionized gas
with overabundant O. The evidence of C and O emission features suggest that the
donor is a carbon/oxygen white dwarf (COWD).  \citet{Bahramian+17} also show
evidence for substantial (factor of $\approx 8$) variations in the X-ray
luminosity on a super-orbital timescale of 6.8 d. These variations have no
straightforward explanation, but could possibly be due to a precessing and/or
warped accretion disk \citep[see e.g.][]{KotzeCharles12}.

The ratio of radio to X-ray luminosity in X9 is most consistent with an
accreting black hole. At the observed X-ray luminosity (a few $\times 10^{33}$
erg s$^{-1}$) accreting neutron stars only rarely show similar radio
luminosities, and these systems are nearly all ``transitional'' millisecond
pulsars that also show unusual, short-term X-ray and radio variability not seen
in X9 \citep[e.g.][]{Bogdanov+15,Deller+15}. Nonetheless, the
phenomenology of accreting neutron stars at these X-ray luminosities is not well
understood. Hence, while the existing data for X9 provide compelling evidence
for an ultra-compact low-mass X-ray binary with a CO white dwarf donor, they do
not definitively distinguish between a black hole or neutron star accretor.

In this paper we assess the theoretical constraints on the nature of 47 Tuc
X9 that arise from its formation.  We analyze the stability of mass transfer
from COWDs onto neutron star and black hole accretors of a range of masses, to
determine which lead to ultra-compact X-ray binaries (UCXBs).  Subsequently we
investigate the likely formation rates of black hole -- white dwarf binaries, to
see whether a black-hole accretor is realistic.

\section{STABILITY ANALYSIS}
\label{sec:observations}

We assume, for the time being, that 47 Tuc X9 is descended from a
dynamically-formed binary containing a carbon-oxygen white dwarf (COWD) and a
compact object, either a neutron star (NS) or a stellar-mass black hole (BH).
Such binaries, if they attain sufficiently small pericenter separations, emit
gravitational radiation and spiral together until the white dwarf fills its
Roche Lobe.  After mass starts to transfer to the compact object there are two
possibilities.  The mass transfer can be unstable; i.e.~take place at an
ever-faster rate until the entire white dwarf is accreted into a disc around the
compact object.  This will lead to a luminous transient, but not a long-lived
UCXB.  Alternatively, the mass-transfer rate may reach a maximum and then
decline again; i.e.~be stable.  The boundary between stability and instability
depends on the interplay between emission of gravitational radiation and the
effects of mass transfer and mass loss from the system.

\citet{Bobrick+17} use a novel smoothed-particle hydrodynamics methodology
to model the highly super-Eddington mass transfer that takes
place in white dwarf -- neutron star binaries.  They find that the majority of
the mass transferred from the white dwarf is lost in a wind from the accretion
disc round the neutron star.  The disc that forms is physically
much larger than the neutron star, with the majority of the mass lying around
the circularisation radius of the incoming material, as expected for
mass-transfer rates much higher than the Eddington rate where viscous processes
are inefficient at returning the angular momentum to the orbit.  Hence the disc
winds are very efficient at removing angular momentum, which accelerates the
mass transfer and cause it to become unstable for all white dwarfs more massive
than about $0.2\,\msun$.

\begin{figure}[tb]
\centering
\includegraphics[width=\linewidth, clip]{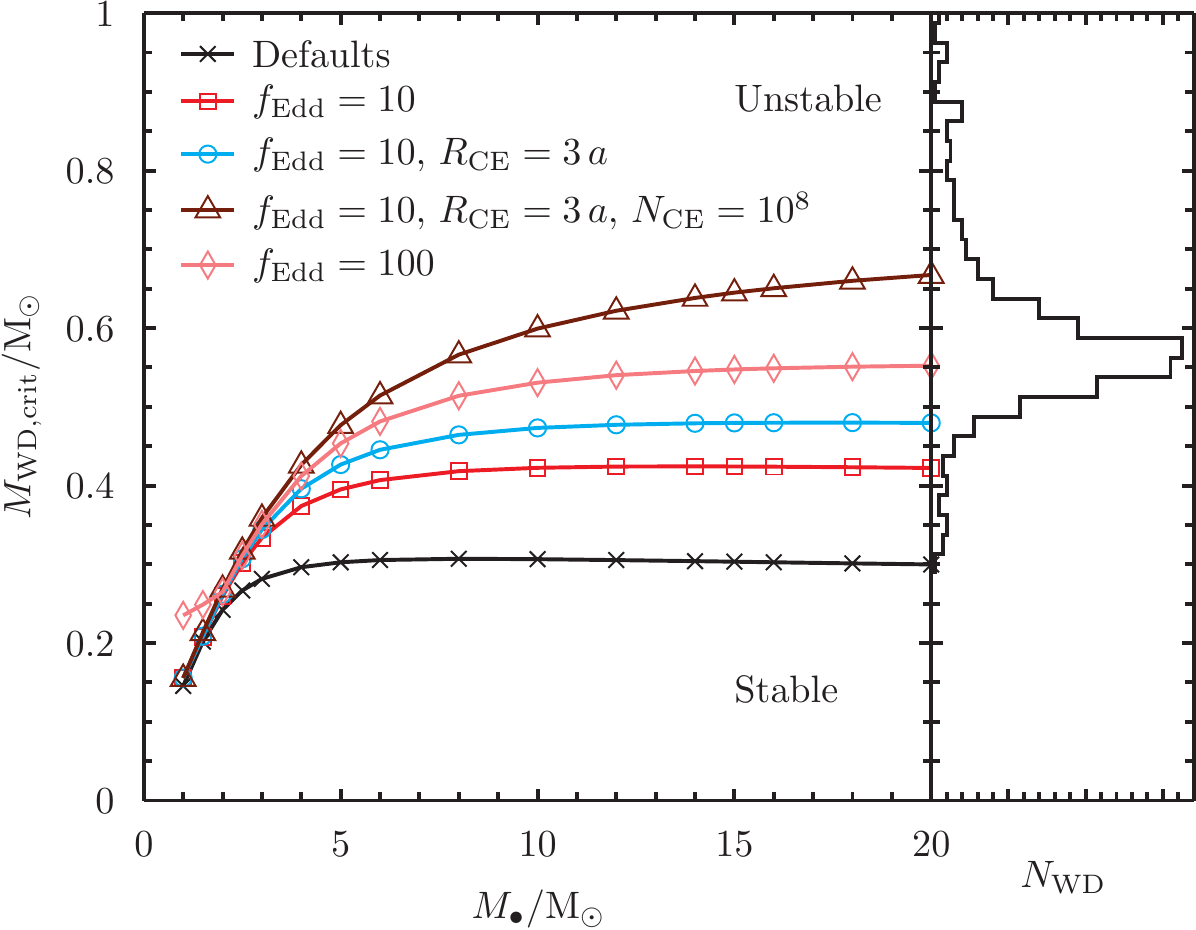}
\caption{
The maximum white dwarf mass, $M_{\rm WD,crit}$, for which stable mass transfer
can take place to a compact object, as a function of compact object mass
$M_\bullet$.  Different lines show different sets of assumptions about the
physics of the binary.  The black line with crosses shows the default
assumptions from \citet{Bobrick+17}, which are as follows.  The maximum
accretion rate is $\dot M_{\rm max}=\dot M_{\rm Edd}$ where $\dot M_{\rm Edd}$
is the the Eddington rate. When high mass-transfer rates lead to 
a common envelope around the binary the mean common-envelope radius is taken to
be $R_{\rm CE}=1.1\,a$ where $a$ is the binary semi-major axis, and the
associated viscous drag timescale is $10^4$ orbital timescales.  Other line
colors and point types show variants of these assumptions.  The red line
(squares) shows $\dot M_{\rm max}=10\,\dot M_{\rm Edd}$.
The blue line (circles) additionally has $R_{\rm CE}=3\,a$.  The brown line
(triangles) additionally has a viscous drag timescale of $10^8$
orbital timescales.  The salmon line (diamonds) has $\dot M_{\rm max}=100\,\dot
M_{\rm Edd}$.  The side panel shows the mass distribution of field white dwarfs
from \citet{DeGennaro+08}. 
}
\label{fig:stab}
\end{figure}

We apply the same model to find $M_{\rm WD,crit}$, the maximum
white dwarf mass for which stable mass transfer takes place, for a range of
compact object masses $M_\bullet$.  Our results are shown in
Figure~\ref{fig:stab}.  The black line with crosses shows the results when adopting
the model of \citet{Bobrick+17} unchanged.  $M_{\rm WD,crit}$ increases with
accretor mass but then flattens off at white dwarf masses of about $0.3\,\msun$.
This would prevent essentially all COWDs from forming UCXBs.  The distribution
of field white dwarf masses from \citet{DeGennaro+08} is shown in the panel to
the right; the peak centered on $\approx 0.55\,\msun$ represents the bulk of
COWDs.   None of these white dwarfs would undergo stable mass transfer under
our standard assumptions.

We have varied the physical assumptions that make up our model to
investigate how they affect the stability.  In particular we vary the maximum
rate at which the compact object can accrete mass, 

\begin{equation}
\dot M_{\rm max,\bullet} = f_{\rm Edd}\dot M_{\rm Edd},
\end{equation}

where $\dot M_{\rm Edd}$ is the Eddington accretion rate.
In \citet{Bobrick+17} we set $f_{\rm Edd}=1$.   We also vary the
size of the common envelope which forms at high mass transfer rates, $R_{\rm
CE}$ as a multiple of the orbital semi-major axis $a$, and the viscous timescale
on which the common envelope extracts energy from the orbit, which we write as a
multiple of the orbital timescale $P_{\rm orb}$; i.e.  $\tau_{\rm
visc,CE}=N_{\rm CE}P_{\rm orb}$.

In our earlier work using $1.4\,\msun$ neutron star accretors we found that the
details of the physical prescription had little effect on the mass-transfer
stability, and that can be seen by the way that the curves in
Figure~\ref{fig:stab} bunch together at low accretor masses.  However, at high
accretor masses the evolution becomes more sensitive to how effectively the
accretion energy drives mass out of the binary.  The assumptions
required for stable mass transfer are on the high side;
accretion of tens of times the Eddington rate, coupled with a loosely-bound
common envelope which is inefficient at dragging on the orbit.

What is clear, however, is that we cannot form a stable system with a neutron
star accreting from a COWD.  Our most optimistic assumptions require the
accretor to be at least six solar masses, which is higher than any model
predicts for the most massive neutron stars \citep{OzelFreire16}.  This implies
that if the donor star in 47 Tuc X9 is a COWD, then the accretor must be a black
hole.

This finding might, at first sight, appear to be inconsistent with reported
observations of NSs accreting from COWDs.  However, where these binaries are in
the field, the donors can equally well be helium stars where all the helium rich
material has already been accreted or lost, leaving just a carbon-oxygen core.
However, for an X-ray binary in a globular cluster, such as 47 Tuc X9, this
scenario is unlikely for two reasons.  Firstly, the progenitors of helium-star
UCXBs are relatively massive, so the UCXB forms less than 2\,Gyr after star
formation \citep{vanHaaften+13}.  Given the age of 47 Tuc this would require an
unfeasibly long-lived UCXB.  Secondly, there is no reason to think that the
He-star UCXB channel would be enhanced in globular clusters.  The only other
UCXB in a globular cluster reported to be have a COWD donor is 4U~0513-40
\citep{Koliopanos+14}.  In that case, however, the authors' diagnosis of a COWD
donor is based on the lack of an iron line, which is in fact likely caused by
the low metallicity in the globular cluster: a He-rich donor fits better with
the observed Type-I X-ray bursts observed from the system (Koliopanos, private
communication).  Hence 47 Tuc X9 is the only binary for which a COWD donor is
{\it required}.  Given this requirement, we go on to consider dynamical
formation of BH--COWD binaries.

\section{DYNAMICAL FORMATION OF BH-COWD BINARIES}
\label{sec:channels}

It is challenging to form close BH-COWD binaries with large mass ratios in the
field because the final stage of mass transfer from the white dwarf's
progenitor is then typically stable owing to the large mass ratio.  This leads
to a wide binary.  Given this, and the interactive dynamical environment
provided by the core of 47 Tuc, it is natural to consider dynamical formation
mechanisms.  The processes relevant to BH-WD binary formation in globular
cluster cores have been described in detail by \citet{Ivanova+10} and we
gratefully adopt their analysis as the foundation of this section.  We first
analyze the expected rates and properties of BH-COWD binaries formed by
collisions of BHs with giant stars,  then briefly discuss the hardening of the
resulting binaries so that they merge.

The core of 47 Tuc has a stellar number density of about $10^5\,{\rm pc}^{-3}$
and a central velocity dispersion of about $12\,{\rm km/s}$
\citep{Meylan89,McLaughlin+06}.  We make the simplifying assumption that these
quantities are constant in time.  All the collisions
that we are interested in are therefore strongly gravitationally focused, 
so the total number of BH-WD binaries forming by direct collisions, $N_{\rm
form}$, is related to the number of black holes in the core, $N_\bullet$, by
$N_{\rm form}=\Gamma N_\bullet$, where

\begin{equation}
\Gamma = 2\pi G f_{\rm p} f_{\rm seg} v_{\infty}^{-1} \sum_i n_i \int_t \left[M_i(t) + M_\bullet\right] R_{\star,i}(t) {\rm d} t,
\label{eqn:ratesum}
\end{equation}

where the sum is over stars of mass $M_i$, radius $R_{\star,i}$ and number
density $n_i$, and the integral runs over the giant phase only.  The velocity at
infinity, $v_\infty$, is taken to be the central velocity dispersion, and
$M_\bullet$ is the mass of the black hole.  Tidal dissipation of
energy in the giant envelope is sufficient to lead to capture of the giant into
a bound binary if the black hole passes within a distance $R_{\rm coll} = f_{\rm
p} R_{\star}$.  \citet{Ivanova+17} model a specific encounter between a black hole
and a red giant in detail using hydrodynamic simulations and find that the maximum
pericenter separation at which tidal capture takes place and the envelope is
completely removed is about three times the giant radius.  Hence we take $f_{\rm
p}=3$.  The factor $f_{\rm seg}$ is an ad-hoc correction for the enhancement of
the number of red giants in the core owing to mass segregation.  We let $f_{\rm
seg}=2$ following \citet{Ivanova+10}.  We evaluate Equation~\ref{eqn:ratesum}
using tracks from \citet{Hurley+00} evaluated at the metallicity of 47 Tuc,
$Z=0.004$ \citep{Harris96}.  The initial mass function is taken from
\citet{Kroupa02} and we take the integral up to the assumed cluster age of
$13.8\,{\rm Gyr}$.

Putting these numbers together, we estimate that about six per cent of black
holes should have had a collision that would lead to a BH-COWD binary (see
Table~\ref{tab:numbers}).  We include collisions with horizontal-branch stars,
which dominate the total numbers.  Once the envelope has been
ejected the resulting object will resemble an sub-dwarf B (sdB) star, which are
produced after an RGB star fills its Roche lobe and the subsquent episode of
common-envelope evolution removes the hydrogen-rich envelope.  We make the
conservative assumption that, once the envelope has been ejected, the core
``burns out'' to form a hybrid white dwarf with a carbon-oxygen center.
More likely is that the resulting sdB star would exhaust core He and form a
helium giant, leading to a second episode of common-envelope evolution and a
shorter merger time.  In either case, when observed as a UCXB, we would expect
such an object to be dominated by carbon and oxygen since the helium-rich outer
layers will be lost during the early mass transfer.  Our numbers are roughly
consistent with \citet{Ivanova+10}; we find a slightly higher rate since we take
the integrated stellar population over time rather than the present-day
population.  In conclusion, if the core of 47 Tuc contains a few tens of
stellar-mass black holes then it is likely that one or more of them has acquired
a degenerate carbon-oxygen companion in a collision with a giant.

\begin{table}[ht]
\centering
\caption{\label{tab:numbers} Total expected BH-WD binaries formed by direct
collisions per BH over 13.8\,Gyr}
\begin{tabular}{lrrr}
\hline
\hline
$M_\bullet/\msun$                     & 5         &     10    &     20    \\
\hline
$N_{\rm form}/N_\bullet$ (all WD types)          & 0.046     &     0.082 &     0.15 \\
$N_{\rm form}/N_\bullet$ (COWD only)            & 0.024     &     0.041 &     0.076 \\
\hline
\end{tabular}
\end{table}

\begin{figure}[tb]
\centering
\includegraphics[width=\linewidth, clip]{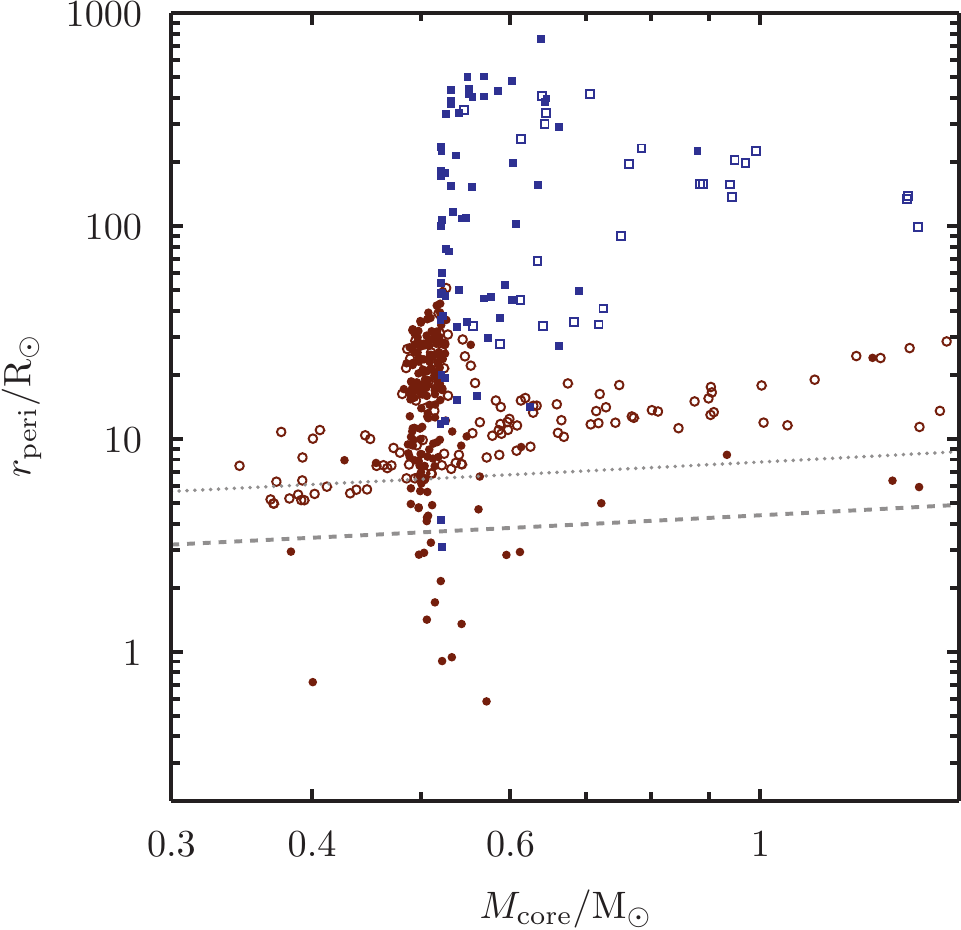}
\caption{
Collisionally-formed binaries containing a $10\,\msun$ black hole and a
carbon-oxygen core.  We plot the binary's pericenter separation, $r_{\rm peri}$,
as a function of core mass $M_{\rm core}$.  Brown circles and blue squares show
binaries formed from collisions with horizontal branch and AGB stars
respectively.  Open symbols indicate circular binaries; filled symbols eccentric
binaries.  Dashed and dotted lines show contours of merger times by
gravitational wave emission for initially circular binaries of $1\,{\rm Gyr}$
and $10\,{\rm Gyr}$.  
}
\label{fig:postCollision}
\end{figure}

By drawing from our predicted collisions uniformly in $\Gamma$ we obtain the
properties expected for the post-collision binaries.
In all cases, since the pericenter separation $R_{\rm coll}\leq
3\,R_\star$, mass transfer will take place at pericenter.  This mass transfer
leads to common-envelope evolution and the shrinking of the binary orbit.
We obtain the final orbital
semi-major axis $a_{\rm f}$ from the standard common-envelope prescription
\citep{Webbink84}:

\begin{equation}
a_{\rm f} = R_\star \frac{\alpha_{\rm
CE}\lambda}{2}\frac{M_\bullet}{M_\star}\frac{M_{\rm \star,core}}{M_{\rm \star,env}}
\end{equation}

where we take the common-envelope efficiency parameter $\alpha_{\rm CE}=1$, and the
structure parameter for the giants $\lambda\approx 1$  as found to
be appropriate for low-mass stars on the horizontal branch \citep{Wang+16}.
The giant's core and envelope masses are $M_{\rm \star,core}$
and $M_{\rm \star,env}$ respectively.  To derive
$r_{\rm peri}$ we consider the separation at closest approach during the initial
encounter, $R_{\rm coll}$.  For gravitationally focused encounters $R_{\rm
coll}$ is uniformly distributed between $0$ and $f_{\rm p}R_\star$.  If $a_{\rm
f}<R_{\rm coll}$ we take the binary to be circular; otherwise we take it to be
eccentric with $r_{\rm peri}=R_{\rm coll}$.  This is a conservative assumption
given that \citet{Ivanova+17} find that the pericenter separation reduces during
common-envelope mass loss.  Figure~\ref{fig:postCollision} shows the pericenter
separations $r_{\rm peri}$ of the binaries that form as a function of core mass.  

It can be seen from Figure~\ref{fig:postCollision} that the typical binary forms
from a collision between a BH and a horizontal branch star with $M_{\rm \star, core}\approx
0.5\,\msun$, and has a pericentre separation between 10 and
$100\,\rsun$. These 
binaries are too wide to merge in a Hubble time by gravitational wave emission 
without some additional help from the cluster.  \citet{Ivanova+10}
argue that they will undergo collisions with wide binaries and form
hierarchical triples.  If the outer orbit is
sufficiently inclined with respect to the inner orbit then Kozai-Lidov cycles
can drive the inner binary up to high eccentricities.  Gravitational wave
emission at pericentre then leads to rapid circularisation and ultimately mass transfer
begins.  They conclude that such triples should form sufficiently frequently that
essentially all of the binaries that we form by collisions will be driven into
mass transfer by Kozai-Lidov cycles.

If the outer binary is too wide, Kozai-Lidov cycles can be suppressed by
relativistic precession of
the inner binary.  Adopting equation 4 of \citet{Hamers+13} to characteristic
values for our problem we obtain the maximum outer orbital semi-major axis
$a_{\rm 2,max}$ that can drive Kozai-Lidov cycles as

\begin{eqnarray}
a_{\rm 2,max} &=& 854\,\rsun
\left(\frac{a_1}{20\,\rsun}\right)^{4/3}\,\left(\frac{M_3}{\msun}\right)^{1/3}\left(\frac{11\,\msun}{M_{\rm \star, core}+M_\bullet}\right)^{2/3} \nonumber\\
              & & \frac{\left(1-e_1^2\right)^{1/3}}{\left(1-e_2^2\right)^{1/2}},
\label{eqn:betacrit}
\end{eqnarray}

where the suffices 1 and 2 denote the inner and outer orbits and $M_3$ is the
mass of the star that orbits the inner binary's center of mass.  For typical
values of $e_1=0.5$ and $e_2=0.9$, $a_{\rm 2,max}=1780\rsun$, which excludes
barely any of the triples that \citet{Ivanova+10} consider.  Hence relativistic
precession should have only a minor impact on this mechanism.

Based on the inferred mass transfer rate of $\approx 10^{-9}\,\msun/{\rm
yr}$ \citep{Miller-Jones+15} and the likely remaining white dwarf mass of a few
per cent of a solar mass, the likely lifetime of 47 Tuc X9's current
state is $10^7$ to $10^8$\,years.  If all the BH-COWD binaries that form over
the lifetime of 47 Tuc merge in the last Gyr -- plausible, given the lengthy
gravitational wave inspiral and Kozai cycles required to merge -- this requires,
on average, tens of binaries to produce a single visible UCXB today.  In turn
that would require a few hundred $10\,\msun$ black holes in the core, which is
at the very upper end of realistic values.  To
derive this pessimistic analysis, however, we have made several conservative
assumptions.  Firstly, when considering the entire Galactic globular cluster
population, 47 Tuc represents only about 2.5 per cent of the total number of
stellar encounters \citep{Bahramian+13}.  Hence, if 47 Tuc X9 is the only
comparable binary in the Galaxy we can divide the required formation rate by a
factor of 40.  Secondly, we neglect binaries that form via exchanges and
subsequent hardening by encounters and triples.  This channel is approximately
equally efficient at producing BH-WD binaries \citep{Ivanova+10}.  We also
ignore the fact that any black holes present are likely to be in binaries with
other black holes.  This increases their cross-section for colliding with giants
as well as potentially providing a built-in companion to drive Kozai cycles.
Given these modifications we argue that formation rates permit a BH-COWD binary.

\begin{figure}[tb]
\centering
\includegraphics[width=\linewidth, clip]{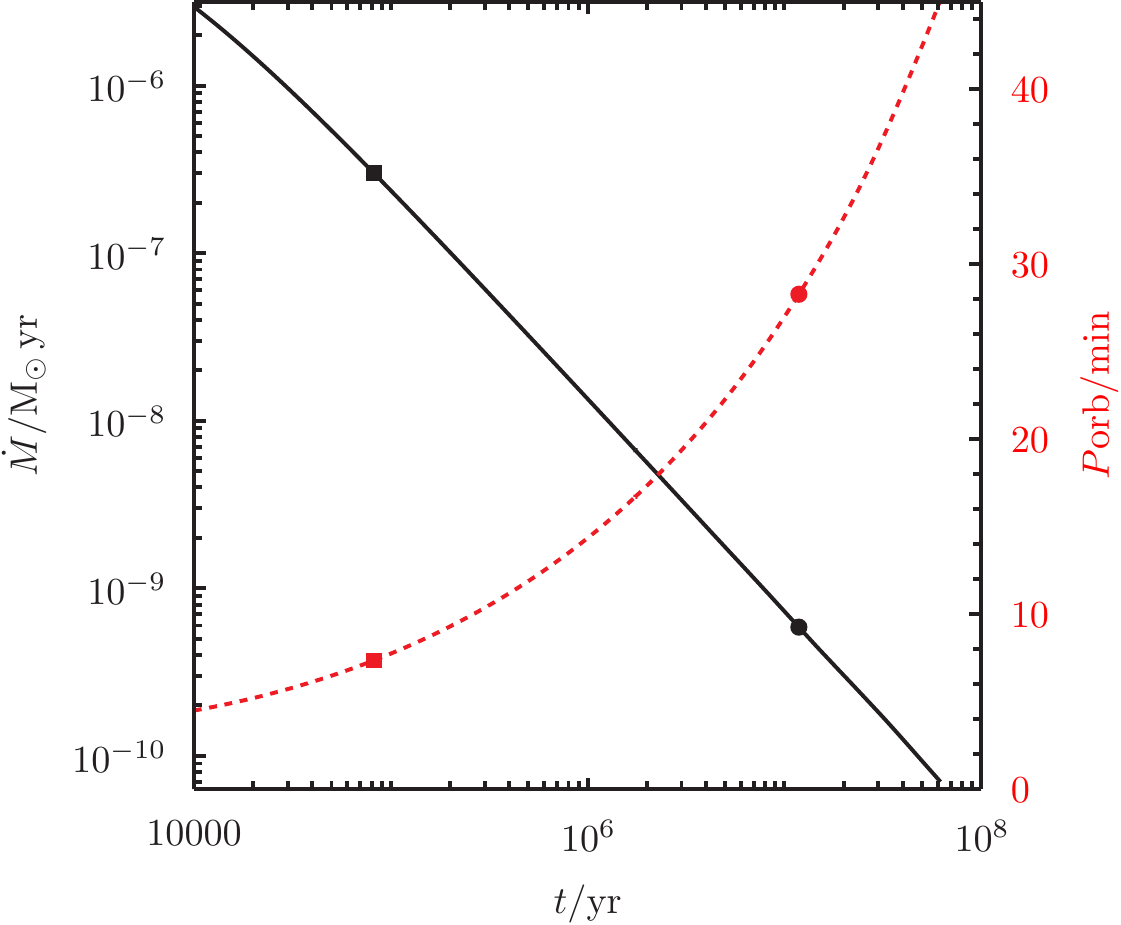}
\caption{
The late-time evolutionary track of a UCXB that forms from
a $0.5\,\msun$ carbon-oxygen white dwarf and a $15\,\msun$ black hole.  The black
line (left-hand ordinate axis) shows the mass transfer rate $\dot M$ as a
function of time since the onset of mass transfer, $t$, in years.  The red
dashed line (right-hand ordinate axis) shows the orbital period $P_{\rm orb}$ in
minutes.  The circles represent the inferred time since the onset of mass
transfer and current mass-transfer rate of 47 Tuc X9, based on an orbital period
of 28.2\,days.  The squares represent the inferred time since the onset of mass
transfer and orbital period of the X-ray source in RZ~2109, assuming a
mass-transfer rate of $3\times10^{-7}\,\msun\,{\rm yr}^{-1}$.
}
\label{fig:MdotPorbEvol}
\end{figure}

A likely extragalactic relative of 47 Tuc X9 is the X-ray source coincident with
RZ~2109, a globular cluster in the Virgo cluster galaxy NGC~4472
\citep{Maccarone+07}.  Its observed properties are consistent
with a stellar-mass black hole accreting from a COWD at around the Eddington
limit \citep{Zepf+07,Zepf+08}.  In our model, such objects form via the same
channel as 47 Tuc X9, but represent an earlier phase of the evolution with
higher $\dot M$.  Figure~\ref{fig:MdotPorbEvol} shows the evolution of the
mass-transfer rate and orbital period of a typical BH-COWD UCXB at late times.
The initial masses of the black and white dwarf are $10\,\msun$ and
$0.5\,\msun$.  The circles and squares indicate the approximate states of 47 Tuc
X9 and the RZ~2109 X-ray source.  Their respective evolution timescales imply
that systems like 47 Tuc X9 should be roughly 150 times as common as systems
like the RZ~2109 X-ray source.  \citet{Peacock+12} and \citet{Caldwell+14}
discuss spectroscopy of about 1600 massive globular clusters in NGC~4472 and M~87
that encompasses a total stellar mass of about $3\times10^9\,\msun$, within
which the only RZ~2109-like object observed was RZ~2109 itself.  47 Tuc X9 is as
of yet unique within the $\approx 2\times10^7\,\msun$ of massive Galactic
globular clusters.  Given the large uncertainty in this rough
comparison, the observed ratio is compatible with our theoretical predictions.


A final question is whether the 6.8 day variability in the 47 Tuc X9's X-ray
flux could be related to the presence of a third star.  If 6.8\,days is the
orbital period of the outer binary it would have semi-major axis $a_{\rm
out}\approx 40\,\rsun$, assuming a $15\,\msun$ black hole.  This is a similar
separation to the dynamically-formed binaries in Figure~\ref{fig:postCollision},
and hence closer than expected to drive Kozai-Lidov cycles,
but possible if the BH--WD binary had been hardened before acquiring this
companion.  Inspired by \citet{Bailyn87} we looked at the eccentricity induced
in the inner binary
by a $1\,\msun$ companion in a moderately eccentric ($e_{\rm out}=0.6$) orbit of
$40\,\rsun$, considering just the gravitational perturbation.  Lidov-Kozai
cycles align the orbital planes at maximum eccentricity, so we took the 
triple to be co-planar.  At each pericenter passage of the outer binary an 
eccentricity of $\approx 10^{-5}$ is excited in the inner binary.
This is sufficient to drive the observed fluctuations in the X-ray emission,
assuming that X-ray
luminosity $L_{\rm X}\propto \dot M$.  In this picture the mass transfer must
damp the eccentricity before the next pericenter passage.  The required damping
timescale is shorter than the viscous timescale of the accretion disc so this is
realistic.  

\section{CONCLUSIONS}
\label{sec:conclusions}

We have explored theoretical constraints on the formation of the luminous X-ray
binary 47 Tuc X9.  Following previous observational work we assume that it is a
compact object (black hole or neutron star) accreting from a carbon-oxygen white
dwarf.  By adapting the model of \citet{Bobrick+17} we have shown that
carbon-oxygen white dwarfs only form stable X-ray binaries when orbiting black
holes of more than $6\,\msun$, and even then only if we allow somewhat
super-Eddington accretion rates and relatively slow extraction of orbital
energy during common-envelope evolution.  Therefore we can rule out a neutron
star as the accretor.  This conclusion is consistent with other
properties of the binary, such as the ratio of its X-ray and radio luminosities.

Analysis of the dynamical formation rates of binaries containing a black hole
and a carbon-oxygen white dwarf suggests that the main channel is the collision
of a stellar-mass black hole with a horizontal-branch star, leading to tidal
capture and, once a bound binary has formed, common-envelope
evolution.  A combination of hardening by Kozai-Lidov cycles in
dynamically-formed triple systems and gravitational wave emission drives such
binaries into contact in less than a Hubble time.  We find that the expected
formation rates are consistent with seeing one such binary in the Milky Way
globular cluster system today. By extrapolation they are also consistent with
the X-ray source in RZ~2109, a globular cluster in the Virgo cluster cluster,
which is in an earlier phase of its evolution.  The observed 6.8-day periodicity
in the X-ray emission from 47 Tuc X9 may be driven by eccentricity induced in
the UCXB's orbit by a perturbing companion.

\acknowledgments

The authors would like to thank Natasha Ivanova, Brian Metzger, and Ben Margalit
for helpful discussions and comments. RPC was supported by the Swedish Research
Council (grant 2012-2254).  RPC and MBD are supported by the project grant
``IMPACT'' from the Knut and Alice Wallenberg Foundation.  JS acknowledges
support from NSF grant AST-1308124, the Packard Foundation, and for HST Program
number GO-14203 from NASA through grant HST-GO-14203.002 from the Space
Telescope Science Institute, which is operated by the    Association of
Universities for Research in Astronomy, Incorporated, under NASA contract
NAS5-26555.  Simulations discussed in this work were performed on resources
provided by the Swedish National Infrastructure for Computing (SNIC) at the
Lunarc cluster, funded in part by the Royal Fysiographic Society of Lund.  The
authors are greatful for the hospitality of the NORDITA programme ``The Physics
of Extreme-Gravity Stars'', during which the ideas discussed in this paper were
conceived.

\end{document}